# Transient Marangoni Waves due to Impulsive Motion of a Submerged Body

**Jian-Jun SHU**

School of Mechanical & Aerospace Engineering, Nanyang Technological University,
50 Nanyang Avenue, Singapore 639798
E-mail: mjjshu@ntu.edu.sg

**Summary**

The Oseen problem in a viscous fluid is formulated for studying the transient free-surface and Marangoni waves generated by the impulsive motion of a submerged body beneath a surface with surfactants. Wave asymptotics and wavefronts for large Reynolds numbers are obtained by employing the Lighthill's two-stage scheme. The results obtained show explicitly the effects of viscosity and surfactants on the Kelvin wakes.

## 1. Introduction

This study is concerned with the transient surface waves generated by the impulsive motion of a submerged body through an incompressible viscous fluid. The surface waves are commonly formed in two distinct types: free-surface waves [1] and Marangoni waves [2]. The free-surface waves mean when the effect of surface tension is insignificant, while the Marangoni waves mean when the effect of surface tension is significant due to contamination by surface-active materials called surfactants.

Lord Kelvin [1] ignored the surface shear stress and developed a theory to determine the kinematics and dynamics of the steady free-surface waves generated by a moving body with constant velocity in an inviscid fluid of infinite depth. He found that the steady free-surface wave pattern consists of a series of so-called diverging waves and transverse waves. The diverging waves spread on each side of the moving body at an acute angle in relation to the body's moving direction, whereas the transverse waves move in the same direction as the moving body. The two wave systems intersect along the so-called "cusp locus" on both sides of the moving body. The angle between this line and the body's moving direction can be calculated as $19^o\,28'$. The lines constitute the outer edge of the so-called Kelvin wake. He also deduced that the diverging waves have a propagation direction of $35^o\,16'$ compared to the moving direction at a certain distance from the body's navigation route. Wehausen & Laitone [3], Chan & Chwang [4] and Shu [5] enriched the theory with the consideration of viscous effect.

In a marine environment, the surface waves are complicated by the contaminant of surfactant due to the Marangoni effect [2], which is the phenomenon of liquid flowing along a surface from places of low surface tension to places of high surface tension. The surfactant concentration varies with the motion of the surface, causing a surface-tension gradient that must be balanced by a non-zero surface shear stress. In this paper, we employ an analytical approach to study the transient interaction of impulsive motion of a submerged body with a contaminated surface, and the consequences in terms of the transient Marangoni waves. We are interested in the effects of viscosity and surfactant on a point force moving beneath a surface covered with a viscoelastic film of negligible thickness. The point force solution can be used to construct solutions for a submerged body of any shape.

## 2. Governing equations

We consider a point force submerged in a viscous incompressible fluid that occupies initially the lower half space $z<0$ in a Cartesian coordinate system. The point force is located at a distance $z_0$ below the surface of the fluid, being suddenly started from rest and made to move with uniform velocity $U^*\vec{e}_x$, where $\vec{e}_x$ denotes the unit vector along the $x$ direction. Let us nondimensionalize the time by $U^*/g$, the distance by $U^{*2}/g$, the velocity by $U^*$ and the pressure by $\rho U^{*2}$, where $g$ is the gravitational constant and $\rho$ the density of the fluid. Because only the wave profile at large distances downstream for high Reynolds numbers is investigated and the surfactant is assumed to act as a linear, viscoelastic film, the dimensionless Navier-Stokes equations describing the fluid flow, induced by an external body force with dimensionless strength $\vec{F}\mathrm{H}(t)\delta(\vec{x}-\vec{x}_0)$, may be linearized as

$$\nabla \bullet \vec{u}^* = 0,\ \frac{\partial \vec{u}^*}{\partial t} + \frac{\partial \vec{u}^*}{\partial x} = -\nabla p^* + \varepsilon \nabla^2 \vec{u}^* + \vec{F}\mathrm{H}(t)\delta(\vec{x}-\vec{x}_0)\,. \tag{1}$$

At negative time $t<0$ everything is at rest, $\vec{u}^* = \vec{O}$, $p^* = 0$, $\xi = 0$, $\zeta = 0$, $\eta = 0$, for $t<0$. On the surface, the tangential stress balances the surface-tension gradient induced by the surfactant and the normal stress has a jump proportional to the surface tension and mean curvature. On $z=0$ we linearize these surface conditions [6] together with the kinematic boundary condition to yield

$$\left.\begin{aligned} \varepsilon\left(\frac{\partial u^*}{\partial z}+\frac{\partial w^*}{\partial x}\right) = \left(\lambda+\kappa\frac{\partial}{\partial t}\right)\left(\frac{\partial^2}{\partial x^2}+\frac{\partial^2}{\partial y^2}\right)\xi,\ & \varepsilon\left(\frac{\partial v^*}{\partial z}+\frac{\partial w^*}{\partial y}\right) = \left(\lambda+\kappa\frac{\partial}{\partial t}\right)\left(\frac{\partial^2}{\partial x^2}+\frac{\partial^2}{\partial y^2}\right)\zeta \\ \eta - p^* + 2\varepsilon\frac{\partial w^*}{\partial z} = \sigma & \left(\frac{\partial^2}{\partial x^2}+\frac{\partial^2}{\partial y^2}\right)\eta \\ \left(\frac{\partial}{\partial t}+\frac{\partial}{\partial x}\right)\xi = u^*,\ \left(\frac{\partial}{\partial t}+\frac{\partial}{\partial x}\right)\zeta = v^*,\ & \left(\frac{\partial}{\partial t}+\frac{\partial}{\partial x}\right)\eta = w^* \end{aligned}\right\}. \tag{2}$$

The fluid velocity and pressure vanish at infinity, $\vec{u}^* \to \vec{O}$, $p^* \to 0$, as $z \to -\infty$. Here the variables $\vec{u}^* = (u^*, v^*, w^*)^T$ and $p^*$ represent the non-dimensional perturbed velocity and perturbed pressure in the fluid, $\xi$, $\zeta$ and $\eta$ are three displacements of the surface along the $x$, $y$ and $z$ directions. $\vec{x} = (x, y, z)^T$ and $\vec{x}_0 = (0, 0, -z_0)^T$ are the field point and source point and the dimensionless parameters $\varepsilon$ and $\sigma$ can be regarded as the reciprocal of the Reynolds number and the Weber number. $\lambda$ and $\kappa$ are the elasticity and viscosity of the viscoelastic film, respectively. $\mathrm{H}(\bullet)$ and $\delta(\bullet)$ are the Heaviside's step function and the Dirac delta function, respectively. The solution to equations (1) for an unbounded fluid is given by Shu & Chwang [7] as

$$\vec{u}_0 = -\frac{\mathrm{H}(t)}{4\pi}\vec{F} \bullet \left(\mathbf{I}\nabla^2 - \nabla\nabla\right)\int_0^t \frac{\operatorname{erf}\left(r^*/2\sqrt{\varepsilon\tau}\right)}{r^*}d\tau,\quad p_0 = \frac{\mathrm{H}(t)\vec{F}\bullet\vec{x}}{4\pi r^3}, \tag{3}$$

where $\vec{x}^* = \vec{x} - \tau\,\vec{e}_x$, $r^* = \left\|\vec{x}^* - \vec{x}_0\right\|$ and $r = \left\|\vec{x} - \vec{x}_0\right\|$. Now let the entire solution be written as $\vec{u}^* = \vec{u}_0 + \vec{u}$, $p^* = p_0 + p$. To reduce the number of variables involved, we represent the motion as a potential flow plus a rotational flow. Thus, we define two new functions $\phi$ and $\vec{\omega} = (\omega_x,\ 0,\ \omega_z)^T$ by

$$\vec{u} = \nabla\phi + \nabla\times\vec{\omega} \tag{4}$$

such that

$$\nabla^2\phi = 0, \quad \frac{\partial\vec{\omega}}{\partial t} + \frac{\partial\vec{\omega}}{\partial x} = \varepsilon\nabla^2\vec{\omega}, \quad p = -\frac{\partial\phi}{\partial t} - \frac{\partial\phi}{\partial x} \tag{5}$$

subject to the conditions $\xi = 0,\ \zeta = 0,\ \eta = 0$, for $t < 0$. Through the Laplace transform in $t$, the Fourier transforms in $x$ and $y$

$$\left[\overline{\phi},\ \overline{\vec{\omega}}\right](s,\alpha,\beta,z) = \int_0^\infty\int_{-\infty}^\infty\int_{-\infty}^\infty [\phi,\vec{\omega}](t,x,y,z)\exp\{-st - i\alpha x - i\beta y - [A,B](z+z_0)\}\,dt\,dx\,dy, \tag{6}$$

$$\left[\overline{\eta},\ \overline{\xi},\ \overline{\zeta}\right](s,\alpha,\beta) = \int_0^\infty\int_{-\infty}^\infty\int_{-\infty}^\infty [\eta,\xi,\ \zeta](t,x,y)\exp(-st - i\alpha x - i\beta y)\,dt\,dx\,dy, \tag{7}$$

then from equation (5), it can be seen that $A$ and $B$ must satisfy

$$A = \sqrt{\alpha^2+\beta^2}, \quad B = \sqrt{\alpha^2+\beta^2+\frac{s+i\alpha}{\varepsilon}}\,. \tag{8}$$

We can express surface condition (2) in terms of $\overline{\phi},\ \overline{\omega}_x,\ \overline{\omega}_z,\ \overline{\eta},\ \overline{\xi}$ and $\overline{\zeta}$ as

$$\mathbf{C}\vec{V} = \vec{C}^{\{A\}}\exp[-A(z+z_0)] + \vec{C}^{\{B\}}\exp[-B(z+z_0)] \tag{9}$$

where the superscripts $\{A\}$ and $\{B\}$ denote the contributions of the potential flow and the rotational flow, respectively, $\vec{V} = \left(\overline{\phi},\ \overline{\omega}_x,\ \overline{\omega}_z,\ \overline{\eta},\ \overline{\xi},\ \overline{\zeta}\right)^T$,

$$\mathbf{C} = \begin{bmatrix} C_{11} & C_{12} & C_{13} & 0 & 0 & 0 \\ C_{21} & C_{22} & C_{23} & 0 & 0 & 0 \\ C_{31} & C_{32} & 0 & 0 & 0 & 0 \\ A & -i\beta & 0 & -(s+i\alpha) & 0 & 0 \\ i\alpha & 0 & i\beta & 0 & -(s+i\alpha) & 0 \\ i\beta\sqrt{\varepsilon} & B\sqrt{\varepsilon} & -i\alpha\sqrt{\varepsilon} & 0 & 0 & -(s+i\alpha)\sqrt{\varepsilon} \end{bmatrix} \tag{10}$$

and

$$\vec{C}^{\{A\}} = \frac{\alpha F_x + iF_z}{2s(s+i\alpha)A}\begin{bmatrix} \alpha A\left[A(\lambda+s\kappa) - 2(s+i\alpha)\varepsilon\right]\sqrt{\varepsilon} \\ \beta A\left[A(\lambda+s\kappa) - 2(s+i\alpha)\varepsilon\right]\sqrt{\varepsilon} \\ i\left[A - (s+i\alpha)^2 + A^3\sigma - 2A^2(s+i\alpha)\varepsilon\right] \\ 0 \\ 0 \\ 0 \end{bmatrix}, \tag{11}$$

$$\vec{C}^{\{B\}} = \frac{F_x}{2s(s+i\alpha)B\sqrt{\varepsilon}} \begin{bmatrix} -A^2(s+i\alpha)(\lambda+s\kappa) \\ 0 \\ -i\alpha\left(1+A^2\sigma\right)B\sqrt{\varepsilon} \\ 0 \\ 0 \\ 0 \end{bmatrix} + O\left(\sqrt{\varepsilon}\right), \tag{12}$$

where

$$C_{11} = i\alpha A\left[A(\lambda+s\kappa)+2(s+i\alpha)\varepsilon\right]\sqrt{\varepsilon}, \quad C_{12} = \alpha\beta(s+i\alpha)\varepsilon^{3/2}, \tag{13}$$

$$C_{13} = i\beta\left[A^2(\lambda+s\kappa)+B(s+i\alpha)\varepsilon\right]\sqrt{\varepsilon}, \quad C_{21} = i\beta A(s+i\alpha)\left[A(\lambda+s\kappa)+2(s+i\alpha)\varepsilon\right]\sqrt{\varepsilon}, \tag{14}$$

$$C_{22} = \left[A^2 B(\lambda+s\kappa)+\left(\beta^2+B^2\right)(s+i\alpha)\varepsilon\right]\sqrt{\varepsilon}, \quad C_{23} = i\alpha\left[A^2(\lambda+s\kappa)+B(s+i\alpha)\varepsilon\right]\sqrt{\varepsilon}, \tag{15}$$

$$C_{31} = (s+i\alpha)^2 + A\left[1+A^2\sigma+2A(s+i\alpha)\varepsilon\right], \quad C_{32} = -i\beta\left[1+A^2\sigma+2B(s+i\alpha)\varepsilon\right]. \tag{16}$$

The solution for the Laplace-Fourier transform of the vertical surface displacement $\eta$ may be expressed as

$$\overline{\eta} = -\frac{i\left(\alpha F_x + iAF_z\right)\left\{(s+i\alpha)^2\left[(s+i\alpha)^2 - A\left(1+A^2\sigma\right)\right] + A^4\left(1+A^2\sigma\right)(\lambda+s\kappa)\varepsilon\right\}\exp\left(-Az_0\right)}{2sA\varDelta} + O\left(\varepsilon^{3/2}\right) \tag{17}$$

where

$$\varDelta(s,\alpha,\beta) = (s+i\alpha)^2\left[(s+i\alpha)^2 + A\left(1+A^2\sigma\right)\right] + A^2\left[4(s+i\alpha)^3 - A^2\left(1+A^2\sigma\right)(\lambda+s\kappa)\right]\varepsilon\ . \tag{18}$$

## 3. Wave asymptotics

The problem posed here is to find the far-field asymptotic behaviour of surface waves induced by the impulsive motion of an immersion point force $\vec{F} = \left(F_x,\ 0,\ F_z\right)$. $F_x$ and $F_z$ represent dimensionless drag and lift forces, respectively. We begin by introducing the cylindrical coordinates $(R,\ \theta)$ on the surface through

$$x = R\cos\theta, \quad y = R\sin\theta\,. \tag{19}$$

To obtain the leading terms in the far-field asymptotic representation for small $\varepsilon$ (large Reynolds number) and small $s$ (large time), we shall employ the Lighthill's two-stage scheme [8], which in essence involves calculating the $\alpha$-integration by residues [9] and the $R$-integration by the method of steepest descent [10]. For the first stage of the Lighthill's scheme, we consider the roots of the pole equation

$$\Delta(s,\alpha,\beta)=(s+i\alpha)^2\left[(s+i\alpha)^2+A\left(1+A^2\sigma\right)\right]+A^2\left[4(s+i\alpha)^3-A^2\left(1+A^2\sigma\right)(\lambda+s\kappa)\right]\varepsilon\,. \tag{20}$$

For small $\varepsilon$ and $s$, the roots $\left(\alpha^{(j)},\ j=1,2\right)$ take the form of

$$\alpha^{(j)}=(-1)^{j-1}A_2-\frac{2iA_1^2 s}{3A_2^2-2A_1(A_1+1)}-\frac{A_1^4\left[4iA_2+(-1)^{j-1}A_1\lambda\right]\varepsilon}{A_2\left[3A_2^2-2A_1(A_1+1)\right]}+O\left(\varepsilon^{3/2}+\varepsilon s+s^2\right), \tag{21}$$

where $A_2=\sqrt{A_1\left(1+A_1^2\sigma\right)}$ and $A_1$ satisfies a cubic equation $\sigma A_1^3-A_1^2+A_1+\beta^2=0$, that is,

$$A_1=\begin{cases}\dfrac{1}{3}\left[1+2\sqrt{1-3\sigma}\cos\left(\dfrac{\pi+\phi}{3}\right)\right], \quad \cos\phi=\dfrac{27\beta^2\sigma^2+9\sigma-2}{2(1-3\sigma)^{3/2}} \qquad 0\le\phi\le\pi & \text{if } \sigma\le\dfrac{1}{4} \\ \text{no real positive root} & \text{if } \sigma>\dfrac{1}{4}\end{cases}. \tag{22}$$

Using the residue theorem of a meromorphic function with respect to $\alpha$, the leading terms contributing significantly to the asymptotic expressions about $\varepsilon=0$ and $s=0$ of the surface elevation can be written as

$$\int_0^\infty \eta\exp(-st)\,dt=-\frac{1}{2\pi s}\sum_{j=1}^{2}\int_{-\infty}^{\infty}\frac{\left[A_2F_x+(-1)^{j-1}iA_1F_z\right]A_1A_2}{3A_2^2-2A_1(A_1+1)}\exp\left(-A_1z_0+iRh_j\right)\left[1+O(\varepsilon+s)\right]d\beta\,, \tag{23}$$

where $h_j\left(\beta\,|\,\alpha^{(j)}\right)=\alpha^{(j)}\cos\theta+\beta\sin\theta$. For the second stage of the Lighthill's scheme, we consider the saddle points that satisfy the derivative of the exponent of the Fourier kernel,

$$\frac{\partial}{\partial\beta}h_j\left(\beta\,|\,\alpha^{(j)}\right)=\frac{\partial\alpha^{(j)}}{\partial\beta}\cos\theta+\sin\theta=0\,. \tag{24}$$

For small $\varepsilon$ and $s$, the roots $\beta_\pm^{(j)}$ take the form of

$$\begin{aligned}\beta_\pm^{(j)}=(-1)^{j-1}\frac{3A_4^2-2(A_3+1)A_3}{3A_4^2-2A_3}&\left\{1+\frac{(-1)^{j-1}4i\left(3A_4^2-4A_3\right)A_4^3 s}{3A_4^6-6A_3^2A_4^4+14A_3A_4^4-28A_3^2A_4^2+8A_3^4+8A_3^3}\right.\\ &\left.+\frac{\left[(-1)^j 8i\left(3A_4^2-4A_3^2\right)A_4^3-\left(3A_4^4-6A_3^2A_4^2+8A_3A_4^2-4A_3^3-4A_3^2\right)A_3\lambda\right]A_3^2\varepsilon}{3A_4^6-6A_3^2A_4^4+14A_3A_4^4-28A_3^2A_4^2+8A_3^4+8A_3^3}\right\}\\ &\times A_4\tan\theta+O\left(\varepsilon^{3/2}+\varepsilon s+s^2\right),\end{aligned} \tag{25}$$

where $A_4=\sqrt{A_3\left(1+A_3^2\sigma\right)}$ and $A_3$ satisfies a sextic equation

$$\left(1+A_3^2\sigma\right)\left(3A_3^2\sigma-2A_3+1\right)^2\tan^2\theta+\left(A_3^2\sigma-A_3+1\right)\left(3A_3^2\sigma+1\right)^2=0\,. \tag{26}$$

The sextic equation (26) cannot be solved by rational operations and root extraction on coefficients. For small $\sigma$, the root $A_3$ takes the form of

$$A_3 = M_\pm\left[1+\frac{(5G_\pm-6)M_\pm^2\sigma}{G_\pm-2}+O(\sigma^2)\right] \tag{27}$$

with the $G_\pm$ and $M_\pm$ written as

$$M_\pm=\frac{G_\pm+1}{2},\quad G_\pm=\frac{1\pm\sqrt{1-8\tan^2\theta}}{4\tan^2\theta}. \tag{28}$$

Hence

$$\begin{aligned} A_1 &= M_\pm+(-1)^j\,\frac{4i(G_\pm-1)M_\pm^{1/2}s}{G_\pm(G_\pm-2)}+\frac{(G_\pm-1)M_\pm^{1/2}\left[(-1)^{j-1}8i(2G_\pm-1)M_\pm^{3/2}+(5G_\pm-2)M_\pm^2\lambda\right]\varepsilon}{G_\pm(G_\pm-2)}\\ &\quad+\frac{(5G_\pm-6)M_\pm^3\sigma}{G_\pm-2}+O\left(\varepsilon^{3/2}+\varepsilon\sigma+\sigma^2+\varepsilon s+\sigma s+s^2\right), \end{aligned} \tag{29}$$

$$\begin{aligned} A_2 &= M_\pm^{1/2}+(-1)^j\,\frac{2i(G_\pm-1)s}{G_\pm(G_\pm-2)}+\frac{(G_\pm-1)M_\pm^{1/2}\left[(-1)^{j-1}8i(2G_\pm-1)M_\pm^{3/2}+(5G_\pm-2)M_\pm^2\lambda\right]\varepsilon}{2G_\pm(G_\pm-2)}\\ &\quad+\frac{(3G_\pm-4)M_\pm^{5/2}\sigma}{G_\pm-2}+O\left(\varepsilon^{3/2}+\varepsilon\sigma+\sigma^2+\varepsilon s+\sigma s+s^2\right), \end{aligned} \tag{30}$$

$$A_4 = M_\pm^{1/2}\left[1+\frac{(3G_\pm-4)M_\pm^2\sigma}{G_\pm-2}+O(\sigma^2)\right], \tag{31}$$

$$\begin{aligned} \alpha_\pm^{(j)} &= (-1)^{j-1}M_\pm^{1/2}\left\{1+(-1)^{j-1}\,\frac{i(G_\pm-3)s}{(G_\pm-2)M_\pm^{1/2}}+\frac{\left[(-1)^{j-1}2i(5G_\pm-7)M_\pm^{3/2}+(3G_\pm-4)M_\pm^2\lambda\right]\varepsilon}{G_\pm-2}\right.\\ &\quad\left.+\frac{(3G_\pm-4)M_\pm^2\sigma}{G_\pm-2}\right\}+O\left(\varepsilon^{3/2}+\varepsilon\sigma+\sigma^2+\varepsilon s+\sigma s+s^2\right), \end{aligned} \tag{32}$$

$$\begin{aligned} \beta_\pm^{(j)} &= (-1)^j G_\pm M_\pm^{1/2}\left\{1+(-1)^j\,\frac{4is}{(G_\pm-2)M_\pm^{1/2}}+\frac{\left[(-1)^{j-1}8i(2G_\pm-1)M_\pm^{3/2}+(5G_\pm-2)M_\pm^2\lambda\right]\varepsilon}{G_\pm-2}\right.\\ &\quad\left.+\frac{(5G_\pm-2)M_\pm^2\sigma}{G_\pm-2}\right\}\tan\theta+O\left(\varepsilon^{3/2}+\varepsilon\sigma+\sigma^2+\varepsilon s+\sigma s+s^2\right). \end{aligned} \tag{33}$$

Using the method of steepest descent, we obtain the leading term

$$\int_0^\infty \eta \exp(-st)\,dt = \frac{1}{s}\sqrt{\frac{1}{2\pi R\cos\theta}}\left(1-8\tan^2\theta\right)^{-1/4}\sum_{j=1}^{2}\sum_{\pm} M_{\pm}^{-1/4}\left[M_{\pm}F_x + (-1)^{j-1} i M_{\pm}^{3/2} F_z\right]$$
$$\times\left[\exp\left(-M_{\pm}z_0 + (-1)^{j-1} iR\frac{M_{\pm}^{3/2}}{G_{\pm}}\left\{1+(-1)^{j-1}\left[4i+(-1)^{j-1}M_{\pm}^{1/2}\lambda\right]M_{\pm}^{3/2}\varepsilon + M_{\pm}^{2}\sigma\right\}\cos\theta \pm (-1)^j i\frac{\pi}{4}\right)\right.$$
$$\left.+\,O\left(\varepsilon+\sigma+s+\frac{1}{R}\right)\right]\exp\left(-\frac{2RM_{\pm}\cos\theta}{\mathrm{G}_{\pm}}s\right). \tag{34}$$

Upon substituting and other mathematical manipulations, the surface elevation can formally be expressed as

$$\int_0^\infty \eta \exp(-st)\,dt = \frac{1}{s}\sqrt{\frac{1}{2\pi R\cos\theta}}\left(1-8\tan^2\theta\right)^{-1/4}\sum_{\pm}\left[M_{\pm}^{3/4}P_{\pm}\left(F_x\cos\gamma_{\pm} - M_{\pm}^{1/2}F_z\sin\gamma_{\pm}\right) + O\left(\varepsilon+\sigma+s+\frac{1}{R}\right)\right]$$
$$\times\exp\left(-\frac{2RM_{\pm}\cos\theta}{\mathrm{G}_{\pm}}s\right) \tag{35}$$

where

$$P_{\pm} = \exp\left(-\frac{M_{\pm}\left(z_0G_{\pm} + 4RM_{\pm}^{2}\varepsilon\cos\theta\right)}{G_{\pm}}\right),\quad \gamma_{\pm} = \frac{R\left[1+M_{\pm}^{2}(\sigma+\lambda\varepsilon)\right]M_{\pm}^{3/2}\cos\theta}{G_{\pm}} \mp \frac{\pi}{4}. \tag{36}$$

Using the inverse Laplace transform with respect to $s$, the exact integral expression of the surface elevation can be written as

$$\eta = \sqrt{\frac{2}{\pi R\cos\theta}}\left(1-8\tan^2\theta\right)^{-1/4}\sum_{\pm} M_{\pm}^{3/4}P_{\pm}\left(F_x\cos\gamma_{\pm} - M_{\pm}^{1/2}F_z\sin\gamma_{\pm}\right)\times \mathrm{H}\left(t-\frac{2RM_{\pm}\cos\theta}{\mathrm{G}_{\pm}}\right) + O\left(\varepsilon+\sigma+\frac{1}{R}\right). \tag{37}$$

## 4. Conclusions

The transient free-surface and Marangoni waves have undergone asymptotic expansion due to the impulsive motion of a submerged point force. The new asymptotic expressions of surface elevations and wavefronts are obtained including the effects of viscosity and surfactants. It is found that the presence of surfactants such as viscoelastic surface films alters the free-surface boundary conditions in the tangential direction and thus strongly modifies the flow pattern. As a consequence, wave energy is dissipated by the enhanced viscous damping in the short-gravity-wave region. So, the presence of viscosity is found to reduce the surface wave amplitude, while the surfactants change the phase of the wave.